\documentclass{aastex}
\begin{document}
\title{\bf Angular separations of the lensed QSO images}
\author{ Pushpa Khare}
\affil{Physics Department, Utkal University,
Bhubaneswar, 751004, India}
\email{khare@iopb.res.in}

\begin{abstract}
We have analyzed the observed image separations of the gravitationally
lensed images of QSOs for a possible correlation with the source
redshift. Contrary to the previously noted anti-correlation based on a
smaller data set, no correlation is found for the currently available
data. We have calculated the average image separations of the lensed
QSOs as a function of source redshifts, for isothermal spheres with
cores in a flat universe, taking into account the amplification bias
caused by lensing. The shape of the distribution of average image
separation as a function of redshift is very robust and is insensitive
to most model parameters. Observations are found to be roughly
consistent with the theoretical results for models which assume the
lens distribution to be (i) Schechter luminosity function which,
however, can not produce images with large separation and (ii) the mass
condensations in a cold dark matter universe, as given by the
Press-Schechter theory if an upper limit of 1-7$\times 10^{13}$
M$\odot$ is assumed on the mass of the condensations.
\end{abstract}
\section{Introduction}
The phenomenon of gravitational lensing is extremely useful for
understanding the large scale structure of the universe.  Studying the
observed properties of the images in individual lenses can provide
knowledge of the mass distribution and mass to light ratio in these
lenses (Grogin and Narayan, 1996; Tyson, Kochaski and Dell'Antonio,
1996).  A statistical analysis of the observed lens systems on the
other hand can help us to restrict the values of the cosmological
parameters (Falco, Kochanek and Munoz, 1998; Link and Pierce, 1998).
Based on the frequency of gravitational lensing, upper limits have been
placed on the value of the cosmological constant (Fukugita et al 1992,
Kochanek, 1996).  It has recently been pointed out by Park and Gott
(1997, hereafter PG) that the image separations of the observed lenses
are strongly negatively correlated with source redshifts, z$_{\rm s}$,
the anti-correlation being much stronger than that predicted by
standard cosmologies.  PG considered several possible ways to
strengthen the anti-correlation.  A steeper mass profile, merger of
galaxies and increase in their mass by in-fall strengthen the
anti-correlation, but not sufficiently to explain the observations.  PG
considered point masses and singular isothermal spheres (SISs) for the
lenses and assumed all the lenses to be identical. They also, did not
consider the effect of amplification bias in their calculation of
average image separation.

Williams (1997, hereafter LLRW) showed that the theoretical upper limit
to the image separations shows a strong anti-correlation with the
source redshift, consistent with the observations, provided (i) the
lensing galaxies have logarithmic surface mass densities that gradually
change with radius, (ii) there is a dispersion in the lensing
properties of galaxies like the central surface mass density or
velocity dispersion and (iii) the characteristic length scale of dark
matter halos of galaxies scales as L$^a$, with a$\simeq$ 0.4.

More data on lensed QSOs have become available since 1997 and the
number of lensed QSOs is now roughly twice the number in the PG
sample.  The PG sample also contained some doubtful lenses. The new
data set has to be examined afresh for the presence or absence of the
correlation between image separation and source redshift. The main aim
of the present paper is to analyze the currently available data to look
for a possible correlation. We then want to compare the observed
distribution of average image separation as a function of source
redshift with the results of a detailed calculation of statistical
lensing of galaxies, with a view of obtaining constraints on the values
of various parameters entering the calculations. The plan of the paper
is as follows. In section 2 we present the analysis of the data using
various statistical tests to quantify the presence of any correlation.
Calculations for lensing statistics in particular of the average image
separation as a function of source redshift, for flat universe, for
realistic lens distributions, taking into account the amplification
bias, for different cosmological models are presented in section 3. In
section 4 and 5 we present the results and conclusions respectively.

\section{Is there a correlation?}

The anti-correlation noted by PG is, as observed by them, mainly due to
five low redshift (z$_s \le$2.15), large separation ($<\Delta\theta>$
greater than 4$\arcsec$) images, namely, 0240-343, 0957+561, 1120+019,
1429-008 and 2345+007. Out of these, only one (0957+561) was a
confirmed lens. Since then Kochanek, Falco and Munoz (1999) have argued
against the lensed nature of most of the wide separation QSO pairs
(hereafter WSQPs). Studying the optical and radio properties of these
pairs they concluded that all of the WSQPs, with $<\Delta\theta>$
between 3 $\arcsec$ and 10 $\arcsec$ are binary quasars rather than
gravitationally lensed images with a one sided 2 $\sigma$ (1 $\sigma$)
upper limit of 22$\%$ (8$\%$) on the lens fraction of these QSO pairs.
They explained the high incidence of occurrence of binary QSOs (which
is 2  orders of magnitude higher than that given by the quasar-quasar
correlation function) as being due to the enhanced quasar activity
during the merger of galaxies. No lensing galaxy could be found in
front of these WSQPs so that if they are lensed images, the lensing
mass, though of the magnitude of a cluster, has to be completely
invisible. Peng et al (1999) have searched for a lensing galaxy in
front of Q1634+267 at the lens optical wavelength using NICMOS and
showed that the lens has to have M/L $\ge$ 690 h$_{65}$ (1200 h$_{65}$)
for $\Omega$ =0.1 (1.0) and H$_0$ = 65 h$_{65}$ km s$^{-1}$
Mpc$^{-1}$.  They therefore, suggest that the double "images" may be
binary QSOs rather than multiple images of a single QSO. Very similar
spectra of the two "images", however, defies this conclusion.  Peng et
al (1999) have compared the spectral similarities of 14 pairs of QSOs
with separations between 3 $\arcsec$ and 10 $\arcsec$ which are having
very similar redshifts (with the velocity differences between pairs
being $\le$ 500 km s$^{-1}$) with the spectral similarities of randomly
chosen QSO pairs from the LBQS.  They conclude that there is $\le 3 \%$
probability that a randomly drawn sample of 14 QSO pairs would show as
similar spectra as the observed pairs. So unless a viable theory to
explain the similarities of the spectra of QSOs in merging galaxies can
be developed, one can not discard the lens hypothesis for the WSQPs.

New data on lensed QSOs have become available since 1997. 49 confirmed
or likely lenses have been compiled by the CfA/Arizona Space Telescope
Lens Survey (CASTLES)\footnote{see http://cfa-www.harvard.edu/castles},
out of which the source redshift is available for 39 QSOs. These
include 15 lenses from the PG sample. For reasons stated above we may
combine these with the 5 WSQPs from the PG sample (0240, 1120, 1429,
1634 and 2345) to get an extended sample of 44 lenses. We have
considered 4 separate samples for our analysis, namely (1) PG20 ; the
sample used by PG of 20 QSOs (2) PG15 ; PG sample without the 5 WSQPs
(3) CAST39; the CASTLES sample of 39 lenses and (4) EXT44; the extended
sample containing 39 CASTLES lenses and 5 WSQPs. The angular separation
vs source redshift for these samples are plotted in Figures 1a, 1b, 1c,
1d respectively.  These samples were searched for the presence of
correlation by performing Spearman rank correlation test. KS test was
also performed to determine if the image separations for sources with
redshift smaller than 2.5 and sources with redshift greater than 2.5
are taken from the same distribution. We also obtained best straight
line fits to the unbinned data of the four samples which are  plotted
in the figure. The results are shown in Table I. It is clear that PG20
shows highly significant anti-correlation between $<\Delta \theta>$ and
z$_s$. The anti-correlation weakens but persists at about 1.8 $\sigma$
level after the removal of the 5 WSQPs (PG15).  CASTLES data on the
other hand do not show any correlation. Adding the 5 WSQPs to CASTLES
data (EXT44) gives rise to anti-correlation but it is very weak and is
statistically insignificant. KS test also shows that for EXT44 sample
the probability that the lens separations for sources with redshift
smaller than and larger than 2.5 are taken from the same distribution
is quite large. We thus conclude that the strong anti-correlation noted
by PG was mainly due to the inclusion of 5 questionable QSOs which
formed a large fraction of about 25\% of their sample. The present data
of 39 confirmed lenses does not show any correlation by itself and even
after the inclusion of the 5 questionable lenses which now form only
11\% of the sample.

\section{Details of calculations} As seen above no correlation exists
between the image separations and the source redshifts and it seems
possible that results of the standard theoretical models may be
consistent with the observations without the need of any drastic
assumptions. It is possible that we may be able to constrain some of
the parameters entering the calculations. In this section we present
the details of our calculation of lensing statistics which differ from
previous calculations only in the way the amplification bias has been
taken into account.

We have assumed the mass distribution of the lenses to be that of ISCs
as found to be required by Maoz and Rix (1993) in order to explain the
lensing events in the HST snapshot survey. We have not taken into
account the effects of ellipticity in the lensing galaxies. As the
ellipticity mainly affects the relative numbers of double and quadruple
lenses (Keeton et al, 1997) it will not affect our results. For the
lens distribution as a function of mass and redshift we consider the
Schechter luminosity function of galaxies and the mass condensations in
a cold dark matter universe obtained using Press-Schechter theory
(hereafter referred to as the PS distribution).  The former is given by
\begin{equation} \Phi (L) dL \propto (L/L^*)^\alpha e^{-L/L^*}
dL,\end{equation} $\Phi$ (L) being the number of galaxies with
luminosity L per unit comoving volume.  Here, $\alpha$ =-1.1 (Marzke et
al, 1998). We have assumed the comoving number density of galaxies to
be independent of redshift. The circular velocity dependence of the
luminosity is taken to be L/L$^{*} = (\sigma/\sigma_{*})^4,
\sigma_{*}=225$ km s$^{-1}$ for elliptical galaxies (Faber \& Jackson,
1976, de Vaucouleurs \& Olson, 1982, Kochanek, 1994) and L/L$^{*} =
(\sigma/\sigma_{*})^{2.6}, \sigma_{*}=144$ km s$^{-1}$  for spiral
galaxies (Tully \& fisher,1977, Fukugita and Turner, 1991), which are
assumed to be 70 $\%$ of all the galaxies (Postman and Geller, 1984).
The PS distribution is given by \begin{equation}\rm n(\rm v_{\rm c},\rm
z)\rm dv_{\rm c}\;=\;{-3(1.67)^{3}\delta_{c}\rm
H_{o}^{3}(1+z)^{5/2}\over{(2\pi)^{3/2}v_{c}^{4}\Delta (r_{o})}}\;{\rm
d\;\rm ln\;\Delta\over{\rm d\;\rm ln\;v_{c}}}\;\times\;\rm
exp({-\delta_{c}^{2}(1+z)^{2}\over{2\;\Delta^{2} (r_{o})}})\;dv_{c}.
\end{equation} Here $\rm n(\rm v_{\rm c},\rm z)$ is the number density
of mass condensations with circular velocity $\rm v_{\rm c}$ at
redshift z, $\delta_{c}$ = 1.68 and the functional form of $\Delta({\rm
r_{o}})$ for the CDM power spectrum of density perturbation is
\begin{equation} \Delta(\rm r_{o})\;=\;16.3\;\rm b ^{-1}\;(1-0.3909\rm
\;r_{o}^{0.1}+0.4814\rm \;r_{o}^{0.2})^{-10}, \end{equation} where b is
the bias parameter.  The mass and circular velocity of a halo are
related to the comoving radius r$_{o}$ and redshift z by,
\begin{equation} \rm M\;=\;{4\;\pi\over{3}}\;\rho_{o}\;\rm
r_{o}^{3},\;\;\;\;\rm V_{c}\;=\;1.67\;(1+\rm z)^{1/2}\;H_{o}\;r_{o},
\end{equation} $\rho_0$ being the mean density of the universe. 
The density fluctuation amplitudes are taken from N-body
simulation work (Narayan and White, 1988, Mo,Miralda-Escude and Rees,
1993).

We consider several flat world models with different values of the
cosmological constant.  Following Fukugita et al (1992), we use the
angular diameter distances between the lens and the observer ($D_{OL}$)
between the source and the observer ($D_{OS}$) and between the lens and
the source ($D_{LS}$) and take the critical impact parameter for SIS
lens to be \begin{equation} a_{cr}= 4 \pi ({\sigma_{||}\over c})^2
{D_{OL} D_{LS}\over D_{OS}},\end{equation} $\sigma_{||}$ being the one
component velocity dispersion for the lens equal to $v_{c}/\sqrt{2}$.
The angular diameter distance formulae for various types of world
models are taken from Fukugita et al (1992). The lensing cross-section
for ISC is given by (Hinshaw and Krauss 1987) $\sigma=\pi l_0^2,\; l_0$
being the maximum impact parameter for lensing, given by
\begin{equation} l_0= [(a_{cr}^2+5a_{cr} r_{c} -0.5r_{c}^2)-0.5
r_{c}^{1/2}(r_{c}+4 a_{cr})^{3/2}]^{1/2},\end{equation} $r_c$ being the
core radius of the lens mass distribution. The amplification of an
image is obtained from $ A={b\over l} {d b\over dl}$. Here $l$ is the
impact parameter and $b$ is the image position in the lens plane
obtained by solving the lens equation\\ \begin{equation} b^3+2 l
b^2+b(l^2+2 a_{cr} r_{c}-a_{cr}^2)+2 l a_{cr} r_{c} =0.\end{equation}
The average image separation for a given value, z$_s$, is given by
\begin{equation}<\Delta \theta>= {{\int\limits_{0}^{z_s}
dz_l\int\limits_{0}^{\infty} dv_c\int\limits_{0}^{l_0} dl\;
{n_s(z_s,z_l,v_c,M_B^{lim})
\Delta\theta(z_s,z_l,v_c,l)}}\over{\int\limits_{0}^{z_s}
dz_l\int\limits_{0}^{\infty} dv_c\int\limits_{0}^{l_0} dl
\;{n_s(z_s,z_l,v_c,M_B^{lim})}}}.\end{equation} Here,
$\Delta\theta(z_s,z_l,v_c,l)$ is the separation between the two
brighter images produced by a lens at z$_l$ with circular velocity
$v_c$ and impact parameter $l$ and $n_s(z_s,z_l,v_c,M_B^{lim})$ is the
number of observable sources (with m$_{obs} < $m$_{lim}$) at the
redshift z$_s$ which will be lensed by galaxies at a redshift z$_l$
with circular velocity $v_c$ and impact parameter $l$, such that both
of the brighter images will have luminosity higher than that for
$M_B^{lim}$. This can be obtained multiplying the QSO luminosity
function by the optical depth for lensing and integrating over the
observable magnitude interval. This can be written as
\begin{equation}n_s(z_s,z_l,v_c,M_B^{lim})={\int\limits_{M_B=-\infty}
^{M_B^{lim}+2.5 log A} dM_B {\phi (M_B)
{\partial^3\tau(z_s,z_l,v_c,l)\over{\partial z_l \partial v_c \partial
l} }}}.\end{equation} Here, M$_B^{lim}$ is the absolute magnitude of a
source at z$_s$ corresponding to the limiting apparent magnitude
m$_B^{lim}$ of the survey, assumed here to be 18. The results are quite
insensitive to the value of m$_B^{lim}$. $A(z_s,z_l,v_c,l)$ is the
amplification of the weaker of the two bright images and $\phi (M_B)$
is the QSO luminosity function, which is taken from Wallington and
Narayan (1993).  $ {\partial^3\tau(z_s,z_l,v_c,l)\over{\partial z_l
\partial v_c \partial l} }$ is the optical depth for lensing of a
source at z$_s$ by lenses with velocity dispersion $v_c$ at redshift
z$_l$ with impact parameter $l$. This is given by \begin{equation}
{\partial^3\tau(z_s,z_l,v_c,l)\over{\partial z_l \partial v_c \partial
l} }=n_l(v_c,z_l) 2 \pi l {c dt\over dz_l},\end{equation}
n$_l(v_c,z_l)$ being the number density of the lenses with circular
velocity $v_c$ at z$_l$ and 2 $\pi\; l\; dl$ being the crosssection for
lensing for impact parameters between $l$ and $dl$.

The unnormalized probability of lensing for a given image separation
$\Delta \theta$ can be computed from \begin {equation} {\partial
p(z_s,\Delta \theta)\over{\partial \Delta\theta}} =
{\int\limits_{0}^{z_s} dz_l \int\limits_{0}^{l_0} dl
m_s(z_s,z_l,M_B^{lim},l,\Delta \theta)},\end{equation} where,
\begin{equation} m_s(z_s,z_l,M_B^{lim},l,\Delta
\theta)={\int\limits_{M_B=-\infty}^{M_B^{lim}+2.5 log A} dM_B {\phi
(M_B) {\partial^3\tau(z_s,z_l,l,\Delta \theta)\over{\partial z_l
\partial \Delta \theta \partial l} }}}\end{equation} is the observable
number of QSOs (obtained as above by multiplying the QSO luminosity
function with the optical depth of lensing and integrating over the
observable magnitude range) lensed by lenses at redshift z$_l$ with
impact parameter $l$ producing image separation $\Delta \theta$,
A(z$_s$,z$_l,l,\Delta \theta$) is the amplification of the weaker of
the two bright images with image separation $\Delta \theta$ and
\begin{equation}{\partial^3\tau(z_s,z_l,l,\Delta \theta)\over{\partial
z_l \partial \Delta \theta \partial l} } = n_l(v_c,z_l) 2 \pi l {dv_c
\over d\Delta \theta} c {dt\over dz_l},\end{equation} $v_c$ being the
circular velocity of lenses at z$_l$ which will yield the image
separation $\Delta \theta$ for impact parameter $l$. For this
calculation we have ignored the dependence of image separation on
impact parameter which is very weak (Hinshaw \& Krauss, 1987) and have
used the value of image separation by the ISCs at zero impact
parameter. We have verified that this assumption does not lead to
errors larger than 1 $\%$. This is given by \begin{equation} \Delta
\theta = 2 {a_{cr} \over D_{OL}} {(1.0-2{r_c\over
a_{cr}})^{1/2}}.\end{equation} The value of $a_{cr}$ for given $\Delta
\theta$ obtained from the above equation is used to obtain the
necessary value of $v_c$ as \begin{equation} v_c=c( {a_{cr} D_{OS}
\over 2 \pi D_{OL} D_{LS}})^{1/2}.\end{equation}.

\section{Results and Discussion} We have plotted in Figure 2a the
unnormalized differential probability $\partial p(z_s,\Delta
\theta)\over{\partial ln(\Delta\theta)}$ as a function of image
separation for the two lens distribution functions for several
redshifts. We have used a value of 0.2 kpc for the core radius for
these calculations.  This is the upper limit obtained by Wallington and
Narayan (1993) from the observed absence of central images for the
lensed QSOs. Models of individual lenses imply somewhat lower core
radii (Kochanek, 1995).  An upper limit of 1.4 kpc has been obtained
from the results of N body simulations of gravitational collapse of
density peaks by Dubinski and Carlberg (1991). The slope of the lens
distribution (as a function of luminosity or circular velocity) for
Schechter luminosity function is steeper than that for the PS
distribution. As a result the probability distribution is broader for
the PS distribution compared to that for the Schechter distribution.
The slope of the probability distribution, for each lens distribution
is, however, almost independent of the source redshift for $\Delta
\theta > 0.4 \arcsec$ which indicates that the average value of
$\Delta \theta $ may not be very sensitive to z$_s$.

Figures 2b and 2c show the effect of varying the values of various
parameters on the differential probability distribution. Figure 2b
shows that increasing the core radius suppresses the probability for
small values of separations (curves 2,3,4,5). However, again, the slope
of the distribution is independent of the value of the core radius for
$\Delta \theta > 0.3 \arcsec$ for r$_c <$ 0.2 kpc. The image separation
increases with the increase in core radius as the probability for small
separations is suppressed (Hinshaw and Krauss, 1987). However, the
effect of including the amplification bias (which increases the
probability for small values of $\Delta \theta$ as can be seen by
comparing curves 3 and 6) cancels the effect of the increase in the
core radius as has been noted by Hinshaw and Krauss (1887). This
results in the probability distribution being almost independent of
r$_c$ for $\Delta \theta > 0.3 \arcsec$ for the assumed range of r$_c$
values. Higher values of core radius change the probability
distribution more significantly for larger values of separations.  The
effect of increasing the bias parameter is to reduce the probability
for large values of $\Delta \theta$ as seen by comparing curve 1 and
3. In Figure 2c we have plotted the probability distribution for three
cosmological models, including the cosmic concordance model proposed by
Ostriker and Steinhaardt (1995). For PS distribution the peak shifts to
lower $\Delta \theta$ values with increase in $\Lambda$.
 
The average image separation as a function of redshift has been plotted
in Figure 3 for several values of parameters.  In this figure we have
also plotted the observed image separations and the best fit straight
lines for the EXT44 sample (small dashed line) and for the CAST39
sample (long dashed line). It is well known that the PS distribution
tends to over predict large separation lenses (Kochanek, 1994; Flores
$\&$ Primack, 1996). As a result the predicted average image separations
are large compared to the observed values. In plotting the results in
Figure 3 we have used a cutoff on the mass of the condensations.  The
probability distribution for a cutoff of 600 km s$^{-1}$ is shown in
Fig2a for comparison.  As is expected, the absence of large masses
reduces the probability for large separations drastically. Porciani et
al (2000) have shown that the probability of lensing is consistent with
observations if one assumes that the condensations with mass  $\ge
3.5\times 10^{13} $M$\odot$ have non-singular mass distribution.
However, as explained above, for r$_c \le$ 0.2 kpc  the effect of
finite core radius is partly compensated by the amplification bias for
$<\Delta \theta> > 0.3\arcsec$ and it is not possible to suppress the
probability for separations $> 4\arcsec$. A cutoff on the mass of
condensation is therefore needed to reduce the probability for these
separations.

Figure 3a shows that the $<\Delta \theta>$ values are almost
independent of the source redshift.  The decrease of $<\Delta \theta>$
with z$_s$ is somewhat more pronounced for the assumption of PS
distribution for small values ($\le$ 2) of z$_s$. The decrease is
larger when the distance formulae for empty beam are used compared to
that for the case of filled beam (curves 1 and 2).  The values of
$<\Delta\theta>$ are sensitive to the values of the upper limit on the
circular velocity of the condensations used in the calculation.  An
increase in the upper limit from 750 km s$^{-1}$ to 1000 km s$^{-1}$
increases the values of $<\Delta \theta>$ by a factor $\ge$1.5 (curves
2 and 4). Change in the bias parameter (curves 4 and 6) again changes
the absolute values but does not alter the slope of the distribution.
As expected from Fig 2b, the values of $<\Delta \theta>$ are almost
independent of the core radius (curves 4 and 5). We thus see that the
observed image separation as a function of redshift (as seen from the
best fit line) is in reasonable agreement (in view of the uncertainties
of the best fit given in Table I) with the theoretical results for an
upper limit on circular velocity of about 600-750 km s$^{-1}$ for b=1
for PS distribution. This value of course is not absolute as higher
values will be needed for higher bias parameters. Use of Schechter
luminosity function, however, produces considerably lower values of
$<\Delta \theta>$.

In Figure 3b we have shown the results for different assumptions about
the change in the core radius with redshift and with circular
velocities of the condensations as suggested by LLRW.  Assuming $r_c$
$\propto {v_c}^{0.5, 1.0, 2.0}$ (curves 4, 5 and 6) does not change the
results significantly.  Similarly the assumption of a redshift
dependence of $r_c$ $\propto (1.0+z_l)^{-1,-2}$ (curves 1 and 2) also
changes the results only by a small amount. This is expected in view of
the very weak dependence of $<\Delta \theta>$ on core radius. Thus the
observed absence of correlation is consistent with the results for a
flat universe even if the core radius varies with luminosity contrary
to the suggestion of LLRW. In Figure 3c we have plotted the results for
different cosmologies. Values of $<\Delta \theta>$ decrease with
increase in $\Lambda$.

The results presented above show that the average image separation is
not very sensitive to (i) the values of core radii (ii) different
assumptions about the dependences of core radii on luminosity or
redshift and (iii) the values of cosmological constant between 0 and
$\simeq$0.65. The values are sensitive to the upper limit used for the
circular velocity for the PS distribution. For b= 1, core radii between
0 and 0.2 kpc and cosmological constant between 0 and $\simeq$ 0.65, an
upper limit of about 600-750 km s$^{-1}$ on the circular velocities of
the dark matter halos for the PS distribution is roughly consistent
with the observations of image separations. It may be noted that the
observed values of image separations do show considerable scatter at
any given source redshift. A large scatter is expected from the flat
probability distributions as a function of $\Delta\theta$ (Figure 2).
The differential probability of lensing as a function of angular
separation can not be directly compared with the observations due to
the small number of observed lenses at any given redshift.  However, if
we take all the observed lenses together, irrespective of their
redshifts, then we see that 6 out of 49 lenses in the full CASTLES data
and 11 out of 54 in the extended data including the WSQPs have image
separation between 3 $\arcsec$ and 8 $\arcsec$. This requires the ratio
of probabilities for separations between 3 $\arcsec$ and 8 $\arcsec$ to
that for separations between 0.3$\arcsec$  and 3 $\arcsec$ to be 0.14
and 0.26 respectively for the two samples. We have calculated this
probability ratio to be 0.62, 0.35 and 0.15 for upper limits of 750,
600 and 500 km s$^{-1}$ respectively for the PS distribution for b=1,
$\Omega$=1 and $r_c$=0.2 kpc. The ratio is 0.04 for the Schechter
function.

Keeton, Christlein and Zabludoff (2000) have considered the detailed
galactic luminosity function dependent upon type and environment
(Bromley 1998a,b). They obtained a good match between their calculated
probability and the image separation distribution for 49 QSO lenses in
the CASTLES data. We have calculated the probability distribution and
the average image separation using the luminosity function used by
them. The results are shown in Figs 1c and 3c. The probability peaks at
a higher value of image separation and as a result the values of
$<\Delta \theta>$ are higher and can be considered to be in agreement
with the CAST39 sample. However we note that even for this luminosity
function it is not possible to get values of image separations larger
than 6 $\arcsec$. The ratio of probabilities for separations between 3
$\arcsec$ and 8 $\arcsec$ to that for separations between 0.3$\arcsec$
and 3 $\arcsec$ is 0.12.
 
\section{Conclusions} We have analyzed the observed image separations
of lensed QSOs for a possible correlation with the source redshift. A
correlation was earlier noted for the then available data by PG. The
present data of 39 confirmed or likely lensed QSOs, even when combined
with 5 wide separation QSO pairs which are doubtful lenses, do not show
any statistically significant correlation. LLRW had shown that if the
core radii scale with luminosity as L$^{0.4}$ then an anti-correlation
is expected theoretically because of the presence of dispersion in
lensing properties of galaxies . We have calculated the average image
separations of the lensed QSOs as a function of source redshifts, for
isothermal spheres with cores in a flat universe, taking into account
the amplification bias caused by lensing. We do not find the strong
anti-correlation stipulated by LLRW. In fact the shape of the
distribution of average image separation as a function of source
redshift is very robust and is insensitive to the change in parameters.
As a result we are unable to obtain meaningful constraints on any of
the parameters. The assumption of the Schechter luminosity function is
unable to produce separations larger than 6 $\arcsec$. The use of PS
distribution on the other hand yields large number of wide separation
lenses even for non-zero core radius and nessecitates the assumption of
a cutoff on the mass of the condensations or large values of core radii
for condensations with large masses. It may be noted that the PS
distribution has also been found to be remarkably successful in
explaining the observed distributions of the QSO absorption lines (Das
\& Khare, 1999).

\acknowledgements

The author is deeply grateful to the referee for his/her extensive,
in-depth and educative comments which have brought the paper to the
present form.
\nopagebreak

\newpage
\figcaption{Image separation vs source redshift for four samples: (a)
PG data of 20 QSOs (b) PG data without the 5 WSQPs (c) CASTLES data of
39 QSOs and (d) CASTLES data with the 5 WSQPs. Triangles represent the
WSQPs. The best fit straight line for each data set is plotted.}
\figcaption{\\  Fig 2a: Unnormalized differential probability of
lensing $\partial p\over \partial ln(\Delta \theta)$ as a function of
angular separation of the bright images. Solid and dashed lines are for
the assumption of PS distribution and Schechter luminosity function
respectively for the lens distribution. Curves labeled 1,2,3,4 and 5
are for source redshifts of 1, 2, 3, 4 and 5 respectively. Dash-dotted
line is for PS distribution, for z$_s =$3 assuming a cutoff of 600 km
s$^{-1}$ on the circular velocities. All curves are for $r_c$=0.2 kpc,
$\Omega$=1.0, b=1 and filled beam.\\
Fig 2b: Unnormalized differential probability of lensing ${\partial
p\over \partial ln(\Delta \theta)}$ as a function of angular separation
of the bright images, for several values of parameters.  Lines labeled
2,3,4,5 are for $r_c$=0, 0.1, 0.2 and 0.5 kpc respectively. Curve 6 has
ignored the amplification bias. All these curves are for b=1. Curve 1
is for b=2 and $r_c$=0.2 kpc.  All curves are for $\Omega$=1, filled
beam and PS distribution.\\
Fig 2c: Unnormalized differential probability of lensing ${\partial
p\over \partial ln(\Delta \theta)}$ as a function of angular separation
of the bright images.  Solid and long dashed lines are for PS
distribution and Schechter luminosity function for the lens
distribution respectively.  Lines labeled 1,2 and 3 are for $\Lambda$=
0, 0.65 and 0.9 respectively.  Short dashed line is for the luminosity
function used by Keeton et al(2000) for $\Lambda=$0.65. All lines,
except for the short dashed line, are for $r_c$=0.2 kpc, b=1 and filled
beam; short dashed line is for $r_c$=0.\label{Fig 1}}
\figcaption{\\ Fig 3a: Average image separation as a function of source
redshift. Curve 8 is for Schechter luminosity function for empty beam.
All other curves are for PS distribution. Curves 1 and 2 are for filled
and empty beams respectively assuming an upper limit of 1000  km
s$^{-1}$ on circular velocity. Curve 3 is obtained by ignoring the
amplification bias. Curve 5 is for SIS lenses. All except curve 6 are
for a bias parameter of 1. Curve 6 is for a bias parameter of 2. All
the curves are for $\Omega$=1.0. Curves 3-6 are for an upper limit on
circular velocity of 750 km s$^{-1}$ while curve 7 is for an upper
limit of 600 km s$^{-1}$. Curves 3-8 are for empty beam and all except
curve 5 have $r_c$=0.2 kpc, curve 5 is for $r_c$=0. The observed image
separations and the best fit straight line (dashed line) for these are
also plotted. The WSQPs are represented by triangles.\\
Fig 3b: Average image separation as a function of source redshift for
PS distribution for different assumptions regarding the core radius.
Curve 1 is for  $r_c \propto (1+z_l)^{-2}$; curve 2 is for $r_c\propto
(1+z_l)^{-1}$; curve 3 is for constant $r_c$; curve 4 is for $r_c
\propto v_c^{0.5}$;  curve 5 is for for $r_c \propto v_c$; and curve 6
is for $r_c \propto v_c^{2}$. All curves are for $\Omega$=1.0, $r_c$
=0.2 kpc (for $z_l$=0 for curves 1 and 2 and for v$_c$=200 for curves
4,5 and 6), b=1, upper limit on circular velocity of 750 km s$^{-1}$
and empty beam.  The observed image separations and the best fit
straight line (dashed line) for these are also plotted. The WSQPs are
represented by triangles.\\
Fig 3c: Average image separation as a function of source redshift.
Curves labeled 1,2 and 3  are for $\Lambda$= 0, 0.65 and 0.9
respectively for PS distribution assuming an upper limit of 750 km
s$^{-1}$ on the circular velocity. Lines marked KE and KF are for the
luminosity function used by Keeton et al (2000) for empty and filled
beams respectively for $r_c$=0. All other lines are for $r_c$=0.2 kpc
and empty beam. All lines are for b=1. The observed image separations
and the best fit straight line (dashed line) for these are also
plotted. The WSQPs are represented by triangles.\label{Fig 3}}
\begin{deluxetable}{ccccccc} 
\tablecolumns{7} 
\tablewidth{0pc} 
\tablecaption{Results of statistical tests and best fit} 
\tablehead{ 
\colhead{Sample}    &  \multicolumn{3}{c}{Spearman rank correlation test} &   \colhead{KS test probability}   & 
\multicolumn{2}{c}{Best fit parameters} \\ 
\cline{2-4} \cline{6-7} \\
\colhead{} & \colhead{$\rho$}   & \colhead{$\sigma$}    & \colhead{$\rho/\sigma$} & 
 \colhead{} & \colhead{a}   & \colhead{b}}
\startdata
PG20 & -0.57 & 0.23 & -2.49 & 1.87(-2) & 6.07$\pm$1.2 & -1.29$\pm$0.47 \\
PG15 & -0.47 & 0.27 & -1.78 & 0.14 & 3.84$\pm$ 1.02 & -0.68$\pm$ 0.37 \\
CAST39 & 6.52(-2) & 0.16 & 0.40 & 0.53 &  2.25$\pm$ 0.60 & -0.13$\pm$ 0.24 \\
EXT44 & -2.56(-2) & 0.15 & -0.17 & 0.39 & 3.23$\pm$ 0.75 & -0.38$\pm$ 0.31 \\ 
\enddata 
\end{deluxetable} 
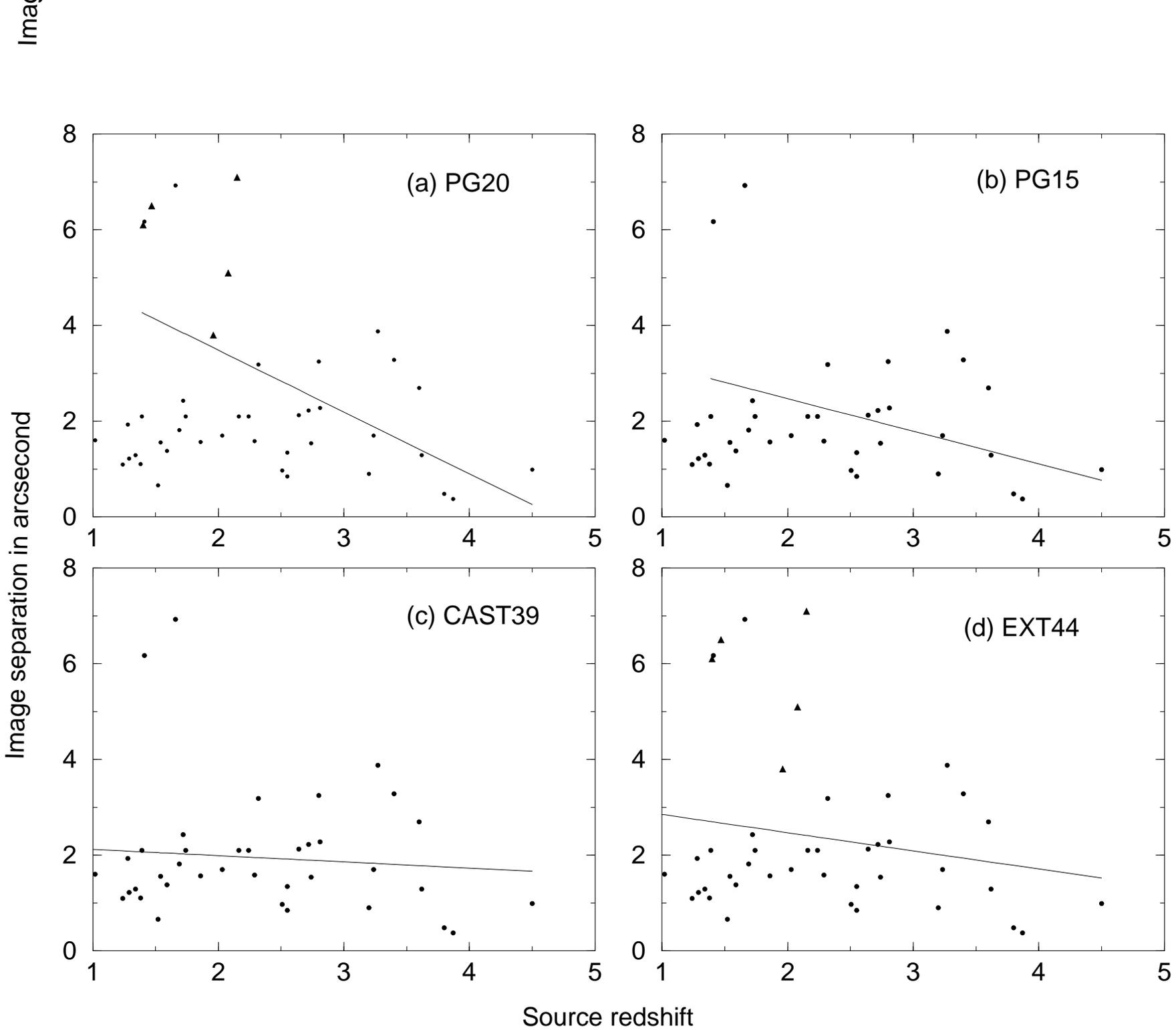
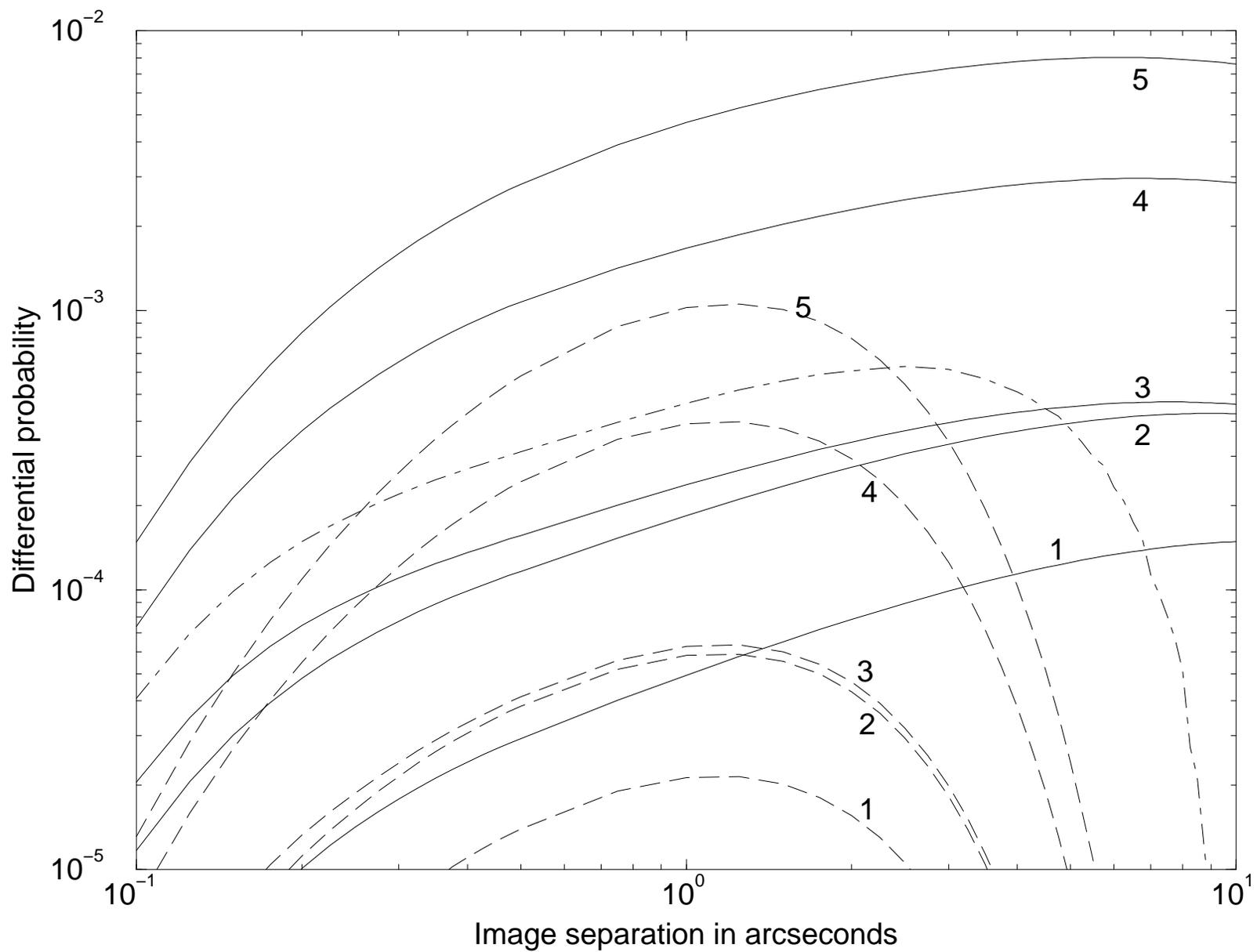
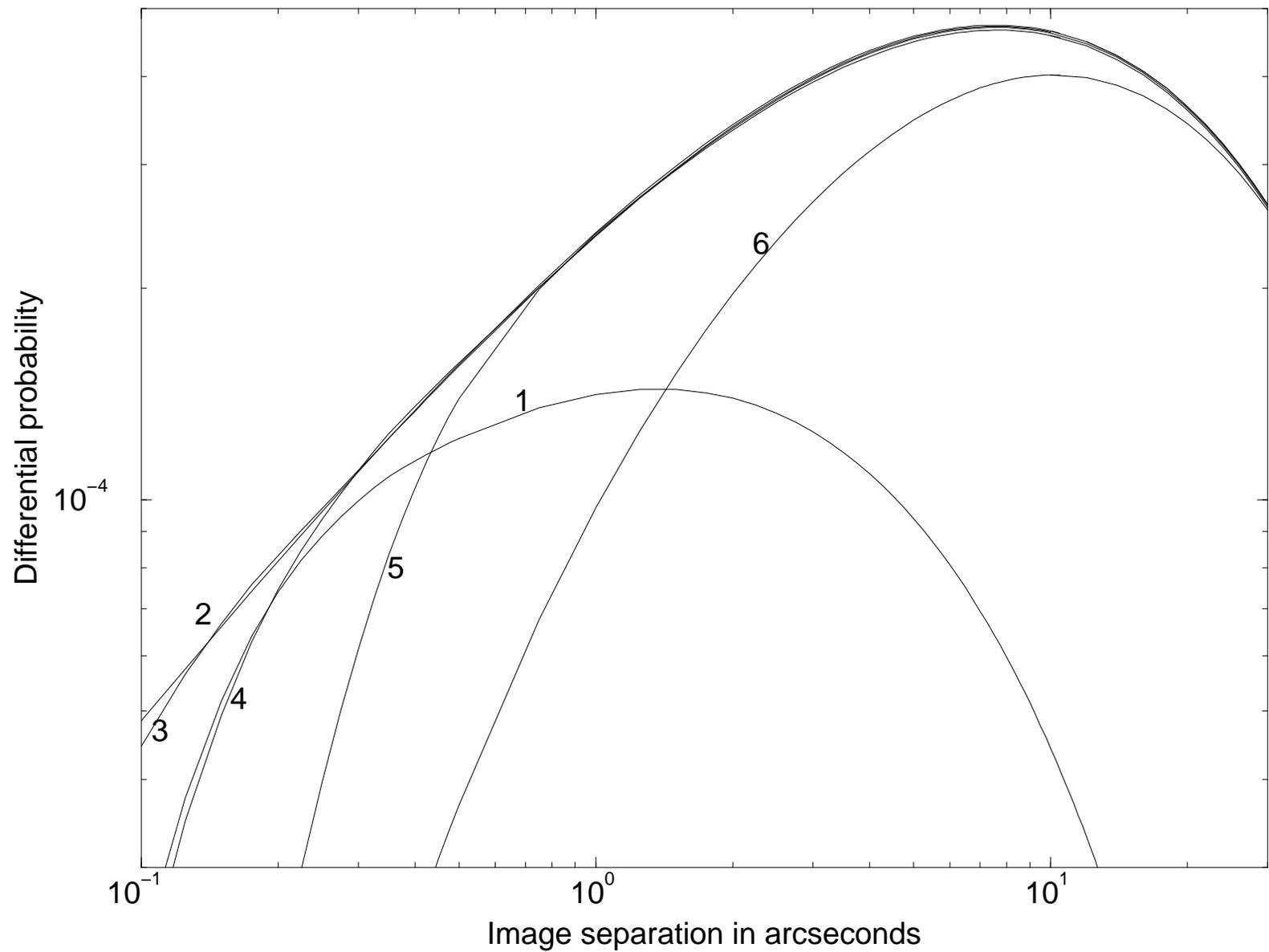
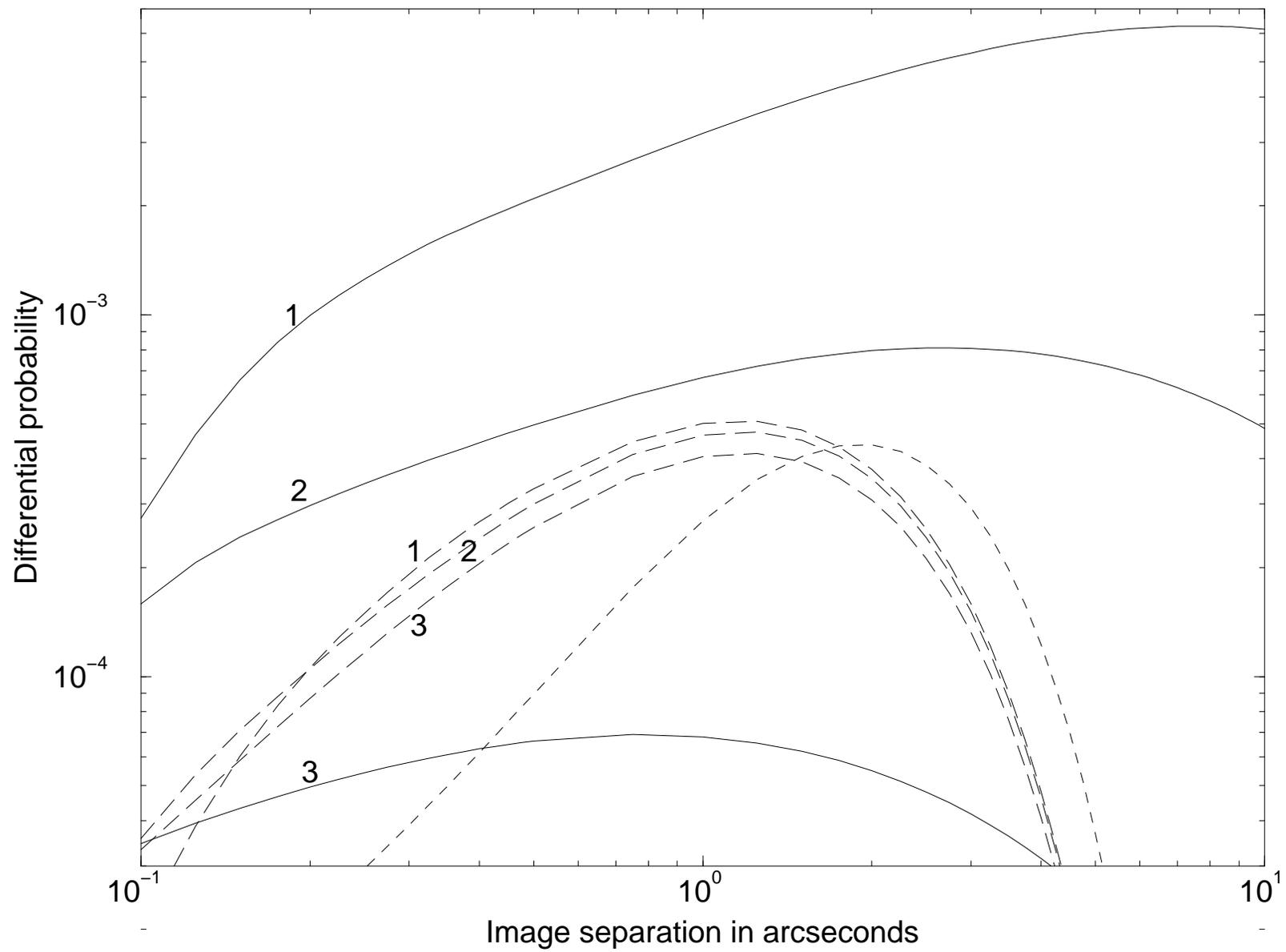
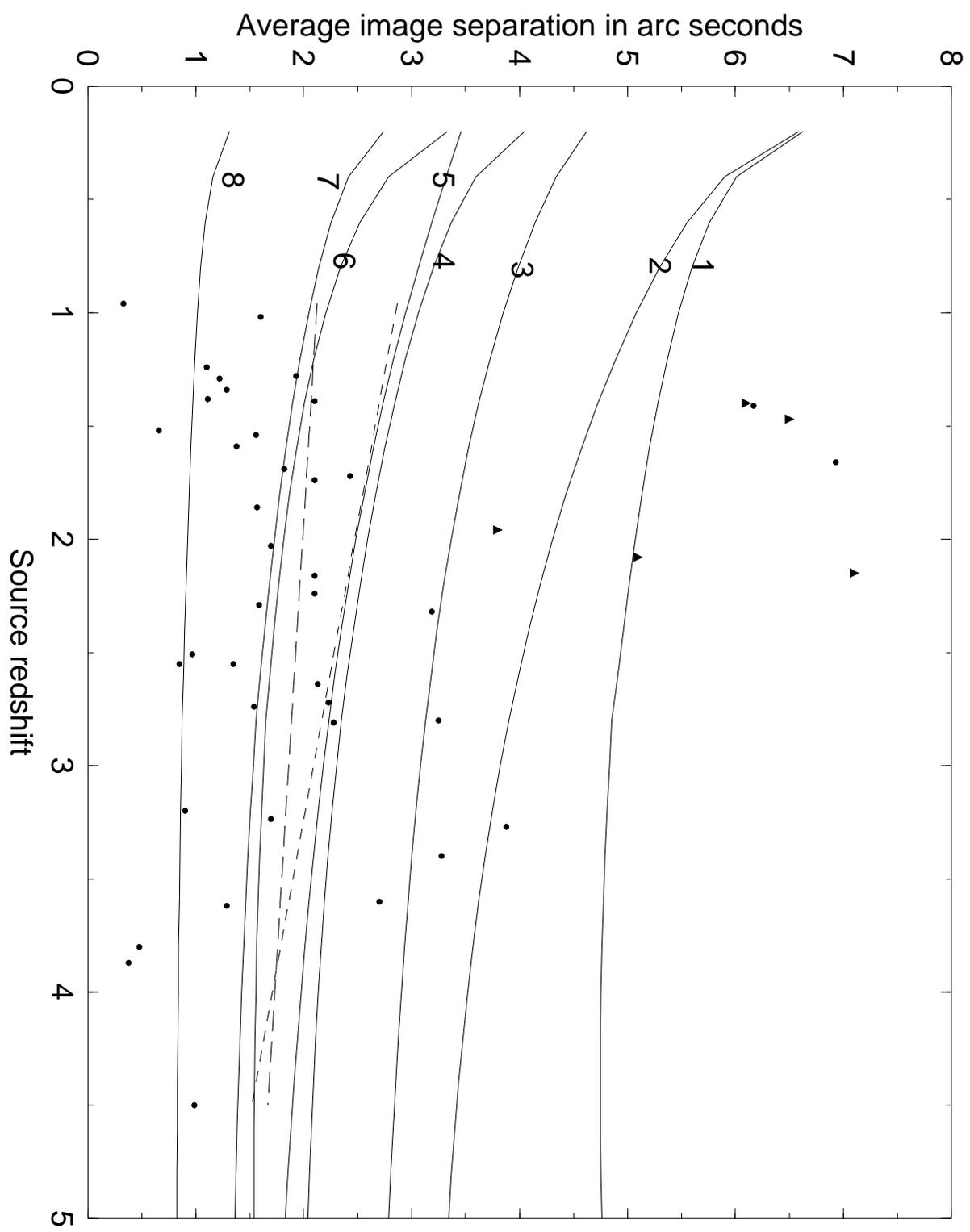
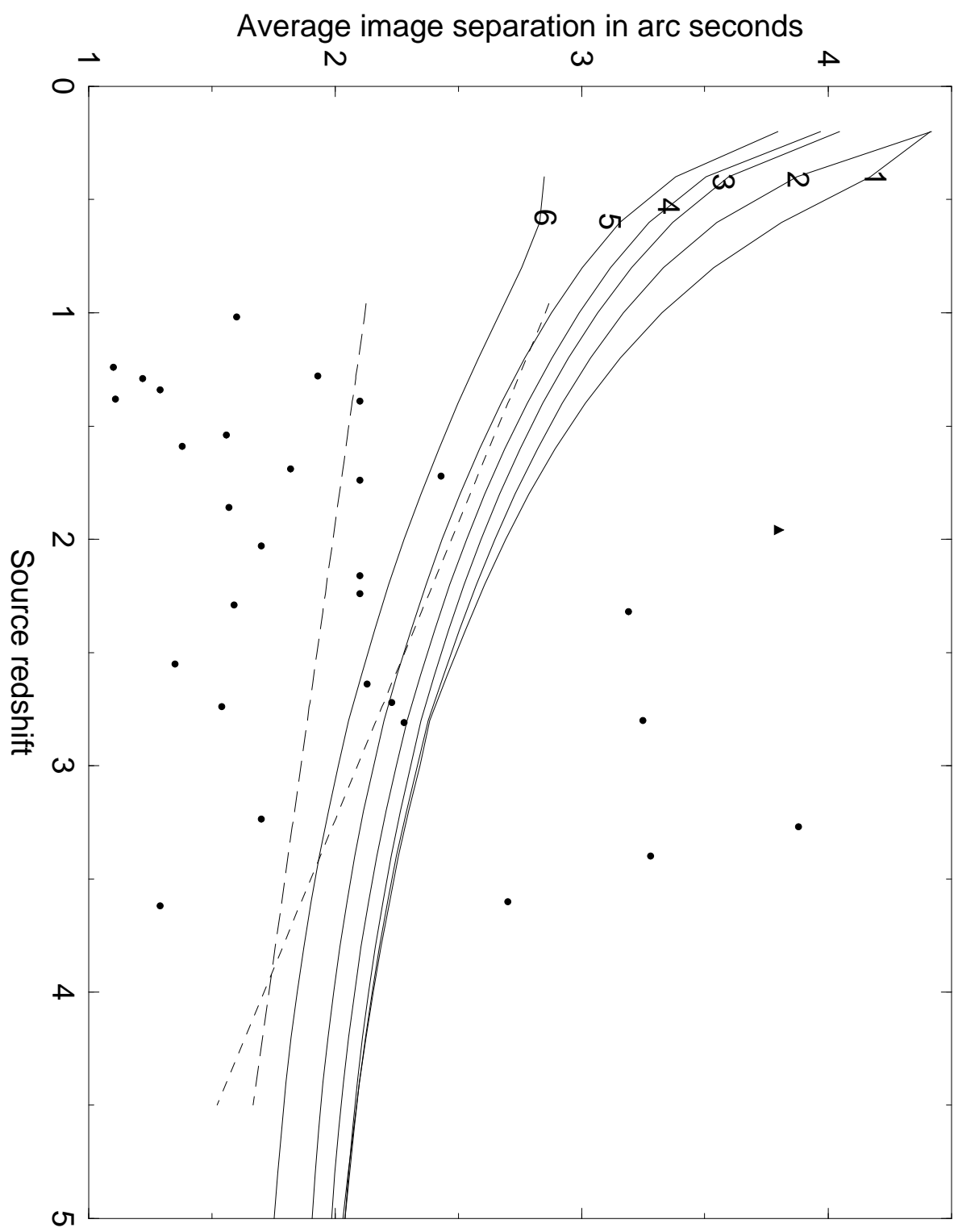
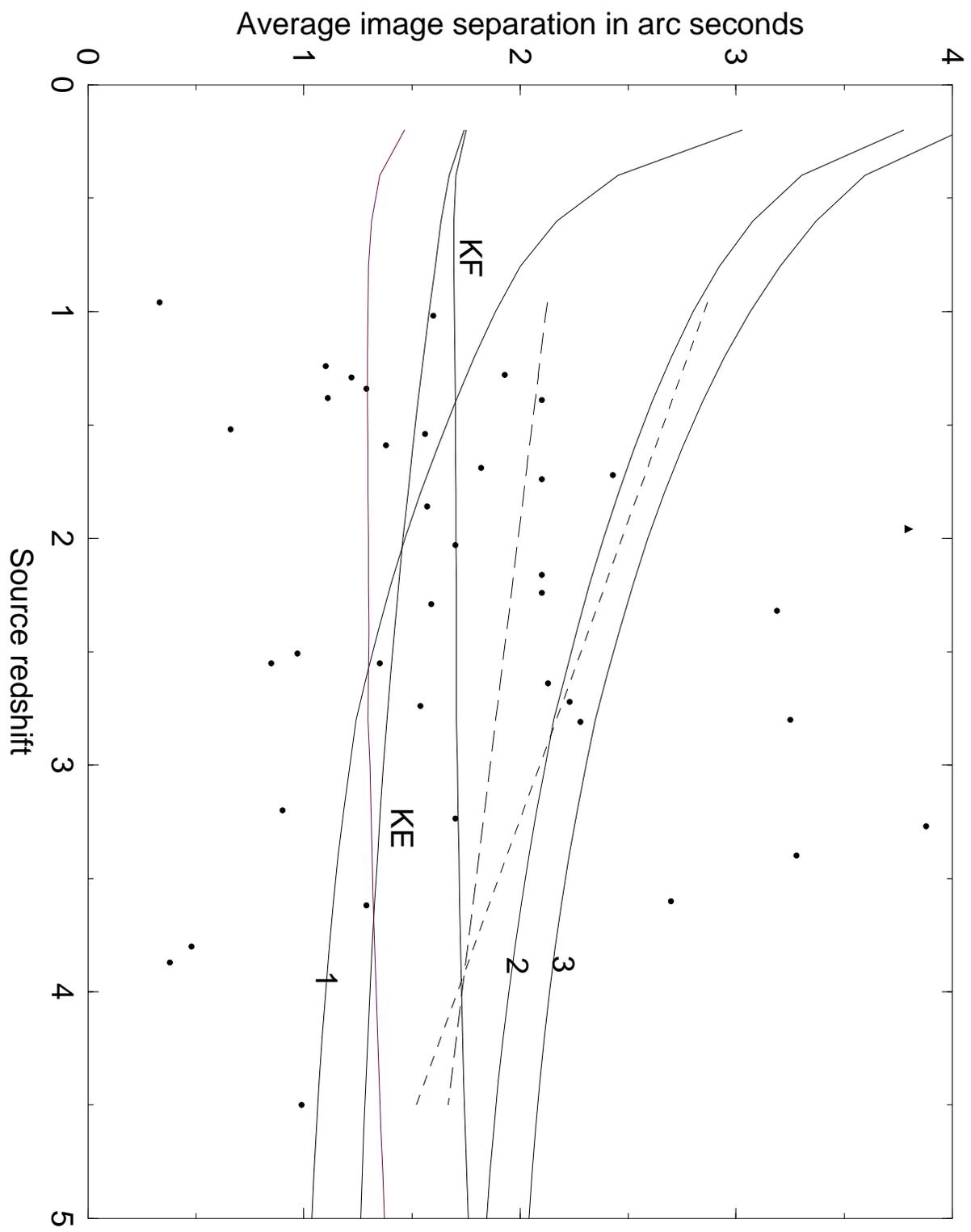
\end{document}